\documentstyle[12pt,epsfig]{article}
\newcommand{\be}[1]{\begin{equation} \label{(#1)}}
\newcommand{\ee}{\end{equation}}
\newcommand{\ba}[1]{\begin{eqnarray} \label{(#1)}}
\newcommand{\ea}{\end{eqnarray}}
\newcommand{\nn}{\nonumber}
\newcommand{\rf}[1]{(\ref{(#1)})}

\begin{document}
\begin{flushright}
\hfill{ CPT-99/P.3925}\\
\hfill{ USM-TH-86}
\end{flushright}
\vskip1cm

\begin{center}
   {\Large\bf CP-violation in high-energy collisions of polarized protons}
\\[3mm]
Sergey Kovalenko$^{a}$\footnote{On leave of absence from
 the Joint Institute for Nuclear Research, Dubna, Russia},
Ivan Schmidt$^{a}$ and Jacques Soffer$^{b}$\\[1mm]
{$^a$\it Departamento de F\'\i sica, Universidad
T\'ecnica Federico Santa Mar\'\i a, Casilla 110-V, Valpara\'\i so, Chile}\\
{$^{b}$\it Centre de Physique Th\'eorique, CNRS Luminy, Case
907, F-13288 Marseille Cedex 09, France}
\end{center}
\bigskip
\begin{abstract}
We study the possibility of a large CP/T-violation in lepton pair
production in the collisions of polarized protons at RHIC-BNL. We
propose the single transverse spin asymmetry which quantifies the
possible CP/T-violating effect in this reaction as a sensitive
indicator of physics beyond the standard model(SM). We consider
two examples of mechanisms beyond the SM which can generate a
non-zero asymmetry of this type. They rely on the R-parity
violating interactions of supersymmetric models and on
phenomenological tensor interactions. We estimate the prospects
for observing this effect in future experiments.
\end{abstract}

\section{Introduction}
CP-violation is one of the most disturbing problem of particle
physics. Being first observed more than 30 years ago in the
neutral kaon system, it has not yet appeared at any other place
except, probably indirectly as baryon asymmetry of the universe,
in the sense that it cannot be understood without CP-violation.

This situation looks puzzling and inspires great efforts both from
theoretical and experimental sides in looking for other CP-violating
phenomena (for a recent review see, for instance, \cite{CP-rev}
and references therein).

In the Standard Model (SM) there is only one source of
CP-violation. This is the single complex phase of the
Cabibbo-Kobayashi-Maskawa (CKM) matrix in the quark sector. Thus,
CP-violation takes place only in the charged current flavor
changing interactions in the quark sector. The physical
CP-violating effect is observable in those processes which involve
at least two CKM matrix elements. In cases in which only one
matrix element is involved, the complex phase can be rotated away
by quark fields redefinition, leaving no physical effect.

Extensions of the SM introduce new
complex phases
enlarging the realm of CP- and T-violation \cite{CP-BSM}. The phenomenological
pattern of this CP/T-violation
%
may significantly differ from the SM picture.
In the SM, for instance, there is no room for CP/T-violation in
$K^0\rightarrow \pi^+ \bar\nu \mu^-$, but once generic scalar
and/or tensor interactions with complex coupling constants are
introduced, CP/T-odd effects become possible in this decay \cite{Leurer}.
The observation of a non-zero transverse muon polarization $P_T^{\mu}$ in
this process, would be a good evidence for the violation of CP-invariance
beyond the SM. Some theoretical estimates have been given \cite{JJGN} and
a new preliminary measurement \cite{E246} indicates
$P_T^{\mu}= -0.0042 \pm 0.0049(stat.) \pm 0.0009(syst.)$,
but the sensitivity is expected to be reduced at the $5 \cdot 10^{-4}$
level.

The best place to look for physics that could exist beyond the SM
is in observables that either vanish or are very suppressed in the
SM. We will see that this is the case for the CP-violating single
spin asymmetry measured in transversely polarized proton-proton
collisions, which furthermore is expected to have very small SM
background. Our purpose is to present this asymmetry as a powerful
indicator for physics beyond the SM, and to estimate its possible
value in some examples of SM extensions. Specifically we consider
the minimal supersymmetric model with R-parity violation and a
phenomenological model with tensor interactions in the quark
sector. Therefore this paper deals with the possibility of
measuring large CP-violation in high-energy collisions, in
processes at a $100 GeV$ scale, an exciting plausible situation
which was already considered a few years ago \cite{IKM}. Earlier,
it was also observed that RHIC might produce CP violating events
in heavy-ion collisions\cite{MS85}.

\section{Single transverse spin asymmetry}
A vast spin physics program will be soon under way at
the RHIC-BNL facility, operating as a polarized $pp$ collider \cite{JS}.
Among the very many spin asymmetries which can be measured, let us
consider the single transverse spin asymmetry
$A_T^{\pm}$ in the reactions
$pp \rightarrow l^+\nu X, l^-\bar\nu X$,
defined as
\begin{equation}
A_T^{\pm}= \frac {d\sigma^{\pm}(\uparrow) - d\sigma^{\pm}(\downarrow)}
{d\sigma^{\pm}(\uparrow) + d\sigma^{\pm}(\downarrow)}~,
\label{A_T}
\end{equation}
where $l=e, \mu, \tau$ and $d\sigma^{\pm}(\uparrow)$
$(d\sigma^{\pm}(\downarrow))$  is the $l^{\pm}\nu$
production cross section with one proton beam
transversely polarized in the {\it up} ({\it down}) direction, with
respect to the normal to the scattering plane. These cross sections
can be computed in the Drell-Yan
picture in terms of the dominant quark-antiquark fusion reactions
\begin{equation}
u+\bar{d}\longrightarrow l ^{+}+\nu ,\;\;\;\;d+\bar{u}\longrightarrow
l
^{-}+\bar{\nu}.  \label{W2}
\end{equation}
For the $l^+\nu$ production an explicit form of the asymmetry in (\ref{A_T})
becomes
\begin{equation}
A_T^{+}= \frac {\langle h_1^u(x_a) \bar d(x_b) \hat{a}_T^u\rangle +
\langle h_1^{\bar d}(x_a) u(x_b) \hat{a}_T^{\bar d}\rangle}
{\langle u(x_a) \bar d(x_b)\rangle  + \langle \bar d(x_a)
u(x_b)\rangle}~.
\label{ATW+}
\end{equation}
The brackets stand for some phase space integrations over the
fractions $x_a$ and $x_b$ of the momenta of the initial protons
($a$ refers to the polarized proton and $b$ to the unpolarized
one), carried by the scattering quark $u$ and antiquark $\bar d$,
whose distributions are $u(x)$ and $\bar d(x)$. Here $h_1^q(x)$
denotes the transversity distribution for quarks (and antiquarks)
\cite{JS1}, which measures the difference of the number of quarks
with transverse polarization parallel and antiparallel to the
proton transverse polarization; and $\hat{a}_T^q$ denotes the
standard single transverse spin asymmetry (polarization or
analyzing power) in the two-body reactions (\ref{W2}), where
$q=u,\bar d$ and $q=d,\bar u$ for the first and the second
reactions respectively. As is known \cite{BLS} this asymmetry can
be expressed in terms of helicity amplitudes
$(\lambda_1,\lambda_2,\lambda_3,\lambda_4)$ of the reaction $\bar
d_{\lambda_1} u_{\lambda_2}\rightarrow \nu_{\lambda_3}
l^+_{\lambda_4}$ where $\lambda_i$ are helicities of the particles
involved in the reaction. There are 5 types of these helicity
amplitudes which we denote as:
$\phi_1^{(i)}=(++,++),(--,--)$;
$\phi_2^{(i)}=(++,--),(--,++)$;
$\phi_3^{(i)}=(+-,+-),(-+,-+)$;
$\phi_4^{(i)}=(+-,-+),(-+,+-)$;
and 8 single-flip amplitudes of
the type $\phi_5^{(i)}=(++,-+),(+-,--), ...$. The expression for
the asymmetry reads
\begin{equation}
\hat{a}_{_T}^{q}=2\ \frac{{\rm Im}\
\sum_{\{i_k\}}(\phi_{1}^{(i_1)}+ \phi_{2}^{(i_2)} +
\phi_{3}^{(i_3)} - \phi_{4}^{(i_4)})^* \sum_i\phi_{5}^{(i_5)}}
{\sum_{\{i_k\}} \left( |\phi_{j}^{(i_1)}|^{2}+|\phi_{j}^{(i_2)}|^{2}+
|\phi_{j}^{(i_3)}|^{2}+ |\phi_{j}^{(i_4)}|^{2} +
2|\phi_{5}^{(i_5)}|^{2}
\right)}.
\label{asym}
\end{equation}
The sum over $\{i_k\}$ implies summation of all the amplitudes of
each class $\phi_j^{(i_k)}$.
Recall that one does not observe the polarization of the final leptons, thus
their helicity states are not fixed.

In the SM the quark subprocesses (\ref{W2}) are mediated by the
s-channel $W^{\pm }$-boson exchange, as shown in Fig.1(a). For the
case of interest  $ \hat{s}>> m_q, m_l$ ($m_q$ and $m_l$ are the
quark and lepton masses), and the corresponding helicity
amplitudes are given by
\ba{SM-ampl}
(+-,--)_l= U_{ud}\frac{g^{2}}{2} \ \frac{m_{l} \sqrt{\hat{s}}}
{\hat{s}-M_{W}^{2}} \sin\theta ,\ \ \
(+-,-+) = -U_{ud} \frac{g^{2}}{2} \
\frac{\hat{s}}{\hat{s}-M_{W}^{2}}(1-\cos\theta ).
\label{f+-SM}
\ea
Notice that only the first amplitude depends on the lepton mass.
Here $g\;$is the $SU(2)_{L}\;$ gauge coupling, $U_{ud}$ is the CKM
matrix element, $\hat s$ and $\theta$ denote the invariant mass of
the lepton pair and the angle between the $u$ quark and the muon
3-momenta in the c.m.s. respectively. As seen from Eq.
(\ref{asym}) these two amplitudes alone do not produce any
contribution to the CP-asymmetry $\hat a_T^q$. We need some source
of a complex phase other than $U_{ud}$. This phase may come from
the new interactions beyond the SM.

Below we discuss two mechanisms which could generate non-zero
CP-asymmetry in Eq. (\ref{asym}). One is suggested by the
supersymmetry (SUSY) and another by a phenomenological model with
tensor interactions in the quark sector.

\section{$R_{p}\hspace{-0.8 em}/\ $ SUSY mechanism}
We start with the minimal R-parity
violating($R_{p}\hspace{-1em}/$ ) SUSY model ($R_{p}\hspace{-1em}/ $ MSSM)
having the superpotential terms
\begin{equation}
W_{R_{p}\hspace{-0.8em}/\;:}=\lambda _{ijk}L_{i}L_{j}E_{k}^{c}+\lambda
_{ijk}^{\prime }L_{i}Q_{j}D_{k}^{c} + \mu_i L_i H_2.  \label{W-Rp}
\end{equation}
in addition to the R-parity conserving terms of the minimal SUSY model(MSSM).
In Eq. (\ref{W-Rp}) $L$, $Q$ stand for lepton and quark $SU(2)_{L}$
doublet left-handed superfields, while $E^{c},D^{c}$ for
lepton and down quark singlet superfields. The coupling constants
$\lambda _{ijk},\ \lambda _{ijk}^{\prime }$ and the mass
parameters $\mu_i$ are in general complex numbers and, therefore,
they can produce CP-violation.
Even if these $R_{p}\hspace{-1em}/\ $-parameters are
taken real, the complex
phases reappear after rephasing the quark and lepton fields, to fix
their masses real and to obtain the standard form for the CKM matrix.
Let us also notice that in the framework of the R-parity conserving MSSM
the only mechanism which can induce the processes
(\ref{W2}) at tree level is
the ordinary W-exchange in Fig.1(a), barring highly suppressed
Higgs-exchange contribution. Thus no new CP-violating phases
are introduced by the MSSM in this process compared to the SM case.

The superpotential (\ref{W-Rp}) gives the following
Lagrangian terms
\begin{eqnarray}
&&{\cal L}_{R_{p}\hspace{-0.8em}/} =
\lambda_{i11}^{\prime}\,\bar{d}P_{L} u\,\,\tilde{e}_{Li}+
\lambda^*_{ij2}\,\bar{\nu}_i P_{R} \mu \,\tilde{e}^*_{Lj}+
\lambda_{i2k}\,\overline{\nu^c}_i P_{L} \mu \,\tilde{e}_{Rk}^*
\nonumber \\
&&+
\lambda_{21k}^{\prime}\,\overline{\mu^c} P_{L} u \,\tilde{d}_{Rk}^*+
\lambda_{i1k}^{\prime *}\, \bar{d} P_{R} \nu^c_i \,\tilde{d}_{Rk}+
\lambda_{ij1}^{\prime}\,\bar{d} P_{L} \nu_i\,\tilde{d}_{Lj}
\label{Lagr}
\end{eqnarray}
generating contributions to
the processes (\ref{W2}) via the s-channel charged slepton
and t-channel down squark exchanges as shown in Fig.1(b,c).
The corresponding helicity amplitudes are
\begin{eqnarray}
\label{Dom}
&&(--,--)_l^{(i)} =
\sum_{m,n,k}
\frac{\lambda^{\prime}_{m11} \lambda^*_{nil} \tilde V_{mk} \tilde
V^*_{nk}}
{\hat s - M^2_{\tilde e_k}} \hat s\  \approx\
\ \eta^{(i)}_{_{l}} \frac{\hat s}{\hat s - M^2_{\tilde e}},\\
\label{t-chan-1}
&&(+-,-+)_l^{(i)}= -\zeta_l^{(i)} \
\frac{\hat{s}/2}{\hat{t}+ M_{\tilde d_L}^{2}}(1-\cos\theta ), \\
\label{t-chan-2}
&&(+-,--)_l^{(i)}= -\zeta_l^{(i)} m_l \
\frac{\sqrt{\hat{s}}/2}{\hat{t}+ M_{\tilde d_L}^{2}}\sin\theta,
\end{eqnarray}
where $\hat{s} = (p_u + p_{\bar d})^2, \ \hat{t} = (p_u - p_l)^2$ with
$p_u, p_{\bar d}, p_l$ being the 4-momenta of the $u,\ \bar d$ quarks and
the final lepton $l$.
The indexes $(i), l$ denote the amplitude for
the final state neutrino $\nu_{i}$ and charged lepton $l$.
We introduced the parameters
\begin{eqnarray}
\label{Dom1}
\eta_l^{(i)} = \sum_{k}
\lambda^{\prime}_{k11} \lambda^*_{kil}, \ \ \
\zeta_l^{(i)}= \sum_{k}
\lambda^{\prime}_{l1k} \lambda^{\prime *}_{i1k}.
\end{eqnarray}
In Eq. (\ref{Dom}) $\tilde V_{mk}$ is the unitary charged slepton
flavor mixing matrix. Analogously the down-squark  flavor mixing
matrix should be introduced in Eqs. (\ref{t-chan-1}),
(\ref{t-chan-2}). Having complex phases, these matrices could
contribute to the CP-asymmetry in Eq. (\ref{asym}). However, we
expect these contributions to be small provided the
slepton(squark) mass spectrum is nearly degenerate $M_{\tilde
e_i}\approx M_{\tilde e}, \ M_{\tilde d_L} \approx M_{\tilde
d_L}$, as suggested by the FCNC constrains. Latter on we neglect
these mixing matrices as indicated by the approximate equality in
(\ref{Dom}). The last possible SUSY source of CP-violation in the
processes (\ref{W2}) is given by the complex phases of the
trilinear soft SUSY breaking parameters $A_{l,d}$ setting the
scale of the Higgs-slepton/squark interaction operators $A_{l}
(H_1 \tilde L \tilde E^c)$, $A_{d} (H_1 \tilde Q \tilde D^c)$. The
physical effect of these phases manifests itself only through the
$(\tilde e_L-\tilde e_R)$ and $(\tilde d_L-\tilde d_R)$ -mixing
and, therefore, is suppressed by the small factor
$m_{e,d}/M_{SUSY}$. Thus in the framework of the
$R_{p}\hspace{-1em}/\ $ MSSM a dominant CP-violating effect in the
processes (\ref{W2}) is expected to originate from the complex
phases $\exp{(-i\phi_{ik}})$ associated with the products of the
$R_p\hspace{-1em}/\ \ $-couplings in Eqs. (\ref{Dom1}).

Substituting (\ref{f+-SM}) and (\ref{Dom})-(\ref{t-chan-2}) into
(\ref{asym}) we obtain in the leading order in small parameters
$\eta,\ \zeta$ the following expression for the CP-asymmetry
\begin{eqnarray}
\label{asym-f} \hat{a}_{_T(l)}^{u}= - \sin\phi_{CP}\ r_l(\hat{s})\
\frac{4\ \sin\theta\ R(\hat{s})\ |\bar\eta_l|} {(1-\cos\theta)^2 +
2 r_l^2(\hat{s}) \sin^2\theta + 4 R^2(\hat{s})\ |\eta_l|^2},
\end{eqnarray}
with
\begin{eqnarray}
\label{def3}
&&\bar\eta_l = \frac{1}{g^2 |U_{ud}|}\sum_{i} \eta_l^{(i)*}, \ \ \
|\eta_l|^2 = \frac{1}{g^4 |U_{ud}|^2}\sum_{i} |\eta_l^{(i)}|^2, \\ \nonumber
&&R(\hat{s}) = - \frac{\hat s - M_W^2}{\hat s - M_{\tilde e}^2}, \ \ \
r_l(\hat{s}) = \frac{m_l}{\sqrt{\hat{s}}}.
\end{eqnarray}
Here the summation runs over all the neutrino flavors $\nu_i$ in
the final state. The angle $\phi_{CP}$ is the relative complex
phase between the $U_{ud}$ and the $R_p\hspace{-1em}/\ \ $ SUSY
parameter $\bar\eta$ in Eq. (\ref{def3}). As seen, the
contribution of the t-channel diagram in Fig. 1(c) drops out from
the leading order expression (\ref{asym-f}).

Given the suppression factor $m_l/\sqrt{\hat{s}}$ the asymmetry
in Eq. (\ref{asym-f}) is larger for the final state $\tau$-lepton
in the process (\ref{W2}). For the main muon channel to be studied at the
RHIC-BNL experiments this asymmetry is smaller by factor $\sim
6\cdot 10^{-2}$.  It is likely that the tau-channel will also be
available in these experiments \cite{tau-BNL}.
First, let us estimate the possible size of the asymmetry
(\ref{asym-f}) for this case. The
CP-asymmetry $\hat{a}_{T(\tau)}^u$ reaches its maximum value
$|\hat{a}_{_T(\tau)}^u| \approx 3.4\cdot \sin\phi_{CP} $ at
$\theta\approx 4.5^o, \ |\bar{\eta_{\tau}}|\cdot R(\hat{s})
\approx  6\cdot 10^{-4}$ being practically independent of the
individual value of the SUSY parameters $\eta$. Of course, any
concrete experiment is limited by the kinematical cut-off
$\theta_{min}\leq \theta \leq \theta_{max},\ 0\leq R(s) \leq
R_{max}$ which can hardly include this point. Therefore, as an
example, let us take a larger angle $\theta = 20^o$. Now the
maximum value of the asymmetry
\begin{eqnarray}
\label{sample1}
|\hat{a}_{_T(\tau)}^u| \approx 0.1 \times \sin{\phi_{CP}}
\end{eqnarray}
is reached at $|\bar{\eta_{\tau}}|\cdot R(\hat{s}) \approx 0.12 $.
In order to get the corresponding values of $\hat{s}$ let us
estimate the value of the parameter $\bar{\eta_{\tau}}$. From the
known constrains on the $R_p\hspace{-1em}/\ \ $ trilinear
couplings $\lambda,\  \lambda'$ \cite{Rp-rev,SK} we can derive the
experimental upper bound $|\bar{\eta_\tau}|\leq 7.3 \times
10^{-2}$. According to the above relation this corresponds to
$\hat{s} \geq 92$GeV, which is not far from the W-pole where
maximal statistics is going to be collected in the RHIC-BNL
experiments. For the same value of the angle, $\theta = 20^o$, the
CP-asymmetry for muon production has the maximum value
\begin{eqnarray}
\label{sample2}
|\hat{a}_{_T(\mu)}^u| \approx 6\times 10^{-3} \ \sin{\phi_{CP}}
\end{eqnarray}
reached at $|\bar{\eta_{\mu}}|\cdot R(\hat{s}) \approx 0.12 $. For
this case we have the upper bound $|\bar{\eta_{\mu}}|\leq
3.0\times 10^{-2}$, derived as in the previous case of tau
production. This corresponds to $\hat{s} \geq 96$ GeV.

The considered mechanism operates out of the W-boson pole since the asymmetry
(\ref{asym-f}) vanishes at $\hat s = M^2_W$. Thus the total event rate
is decreased compared to possible mechanisms operating at the W-pole.

\section{Tensor mechanism}
Let us study an example of the W-pole mechanism in the effective theory
with the Lagrangian term \cite{KLY}
\ba{pole-mech} {\cal L}_T = \frac{1}{\Lambda}
\partial_{\nu}W^-_{\mu} \bar d\ \sigma^{\mu\nu}(f^L P_L + f^R
P_R)u + ..., \ea which may appear at low energies as a footprint
of new physics at the scale $\Lambda$.
Here $\sigma^{\mu\nu}=(i/2)[\gamma^{\mu},\gamma^{\nu}]$.
The lepton sector contains
only the SM interactions. A contribution of the operator
\rf{pole-mech} to the processes (\ref{W2}) is described by
the same diagram Fig. 1(a) as in the SM case. The corresponding
amplitude is given at $\sqrt{\hat{s}}>>m_l,\ m_q$ by the expression
\ba{tensor}
 (++,-+) = \frac{g_2}{\Lambda}
\frac{\sin\theta}{\hat s -M_W^2}
\frac{\hat s^{3/2}}{\sqrt{2}} \left(|f_L|^2 +|f_R|^2\right)^{1/2}
\ea
which does not depend on the type of the final lepton pair
$l \nu$ in the reaction (\ref{W2}).
Now we can combine this amplitude with the SM amplitude $(+-,-+)$
in Eq. (\ref{f+-SM}) to calculate a dominant contribution to the
asymmetry $\hat a_T^u$. The result is
\ba{asym-tens} \hat a_T^u = -2 \frac{\Lambda}{M_W}\ \frac{c\
r(\theta)\ {\rm Im}[U^*_{ud}(f_L + f_R)]}{|U_{ud}|^2 c^2
r^2(\theta)(\Lambda/M_W)^2 +2(|f_L|^2 +|f_R|^2)}, \ea where
$c=(2\sqrt{2} G_F)^{1/2}M_W$, $r({\theta})= \tan^{-1}(\theta/2)$.
In order to make an estimate of the possible size of this W-pole
asymmetry we assume that $f_L\sim f_R\sim f$. Then ${\rm
Im}[U^*_{ud}(f_L +f_R)]\approx 2 |U_{ud}||f|\sin\phi$, where
$\phi$ is a relative complex phase of $U^*_{ud}$ and $f$. This is
a physical phase which can not be removed by the quark and lepton
field redefinition. With this approximation we find that the
asymmetry in Eq. \rf{asym-tens} takes its maximal value
\ba{mas-T} |\hat a_T^u|_{max} = \sin\phi\leq 1, \ \ \ \mbox{at} \
\ \ \tan(\theta/2) = 0.3 \left(\frac{\Lambda}{f M_W}\right). \ea
Notice that for $\Lambda/(f M_W)\rightarrow \infty$ the asymmetry
reaches its maximal value at the scattering angle
$\theta\rightarrow \pi$. If we take $\Lambda \sim M_W, \ |f|\sim
1$ then the asymmetry is maximal at $\theta = 17^o$.

\section{Proton level asymmetry}

Now we can estimate the possible size of the proton level
asymmetry $A_T^{\pm}$ given by formula (\ref{ATW+}). This
quantity, as we argued at the beginning, can be measured in
polarized proton-proton leptoproduction. For estimation purposes
we can significantly simplify formula (\ref{ATW+}). The second
term in the numerator of (\ref{ATW+}) is certainly small, since
one believes that $h_1^u \gg h_1^{\bar d}$ (see ref.\cite{JS1}).
On the other hand, if we select a region where the first term in
the denominator dominates (i.e. $x_a \gg x_b$), we end up with
\begin{eqnarray}
\label{CP-hadron}
\langle A_T^{+}\rangle \approx \langle\frac{h_1^u}{u}\rangle\cdot
\hat{a}_T^u~.
\end{eqnarray}
Using some reasonable theoretical estimates (see ref.\cite{JS1}) one has
$\langle h_1^u/u\rangle  \approx 0.5$. Therefore we finally
obtain
\begin{eqnarray}
\nn
R_{p}\hspace{-1em}/\ \  SUSY:\ &pp\rightarrow \tau\nu X&\ \ \
|\langle A_T^{+}\rangle| \leq 5\times 10^{-2}
\ \ \mbox{at} \ \ \ \theta = 20^o, \hat{s}=92\mbox{GeV}\\ \nn
&pp\rightarrow \mu\nu X&\ \ \
|\langle A_T^{+}\rangle| \leq 3\times 10^{-3} \ \ \mbox{at} \ \ \
\theta = 20^o, \hat{s}=96\mbox{GeV}\\  \nn
\mbox{Tensors}:\ &\ pp\rightarrow l^+\nu X&\ \ \
|\langle A_T^{+}\rangle| \leq 0.5 \ \ \  \mbox{at} \ \ \
\tan{(\theta/2)} = 0.3 \left(\frac{\Lambda}{f M_W}\right).
\end{eqnarray}
These constraints represent certain kinematical samples described
in the previous sections  which aimed to demonstrate possible size of
the CP-asymmetry $|\langle A_T^{+}\rangle|$. Fore a concrete experiment
one has to analyze on the basis of
Eqs. (\ref{asym-f})-\rf{asym-tens} the whole accessible kinematical
domain to find out preferable scattering angles and to
determine a searching strategy.

The SM backgrounds are expected to be small, due to the fact that the
W-production vertex involves only one CKM matrix element even at
the one-loop level, and in order for the asymmetry to be
non-vanishing, several CKM matrix elements, whose relative
convention-independent phase is non-zero, have to be present.
Higher twist contributions are also expected to be small because they
should be suppressed, at least by a factor $M_{proton}/M_{W}$, in addition
to a possible cancellation between the vector and the axial vertices.

\section{Conclusions}
Summarizing, we have shown that there exists the exciting
possibility of searching for the CP violation in polarized p-p collisions.
We proposed an appropriate  CP-asymmetry $A_T$ defined in
(\ref{A_T}) and demonstrated that it may serve as a powerful tool
in order to probe new physics beyond the SM. If at a given
sensitivity the experiment does not observe the CP-odd effect,
setting an upper bound on the asymmetry $A_T$, this will still
allow one to establish new important upper bounds on the
parameters of new physics.

\section*{Acknowledgments}

One of us (JS) would like to thank V.~L. Rykov and F.~M. Renard
for some useful correspondence at  the earlier stage of this work,
and C.~P.~Yuan for several enlightening discussions. This work was
supported in part by Fondecyt (Chile) under grants 1990806,
1000717, 1980150 and 8000017, and by a C\'atedra Presidencial
(Chile).
%

\newpage
%
%
\begin{figure}[htb]
\vspace{1.cm} \hspace{1.0cm} \mbox{\epsfxsize=14.
cm
\epsffile{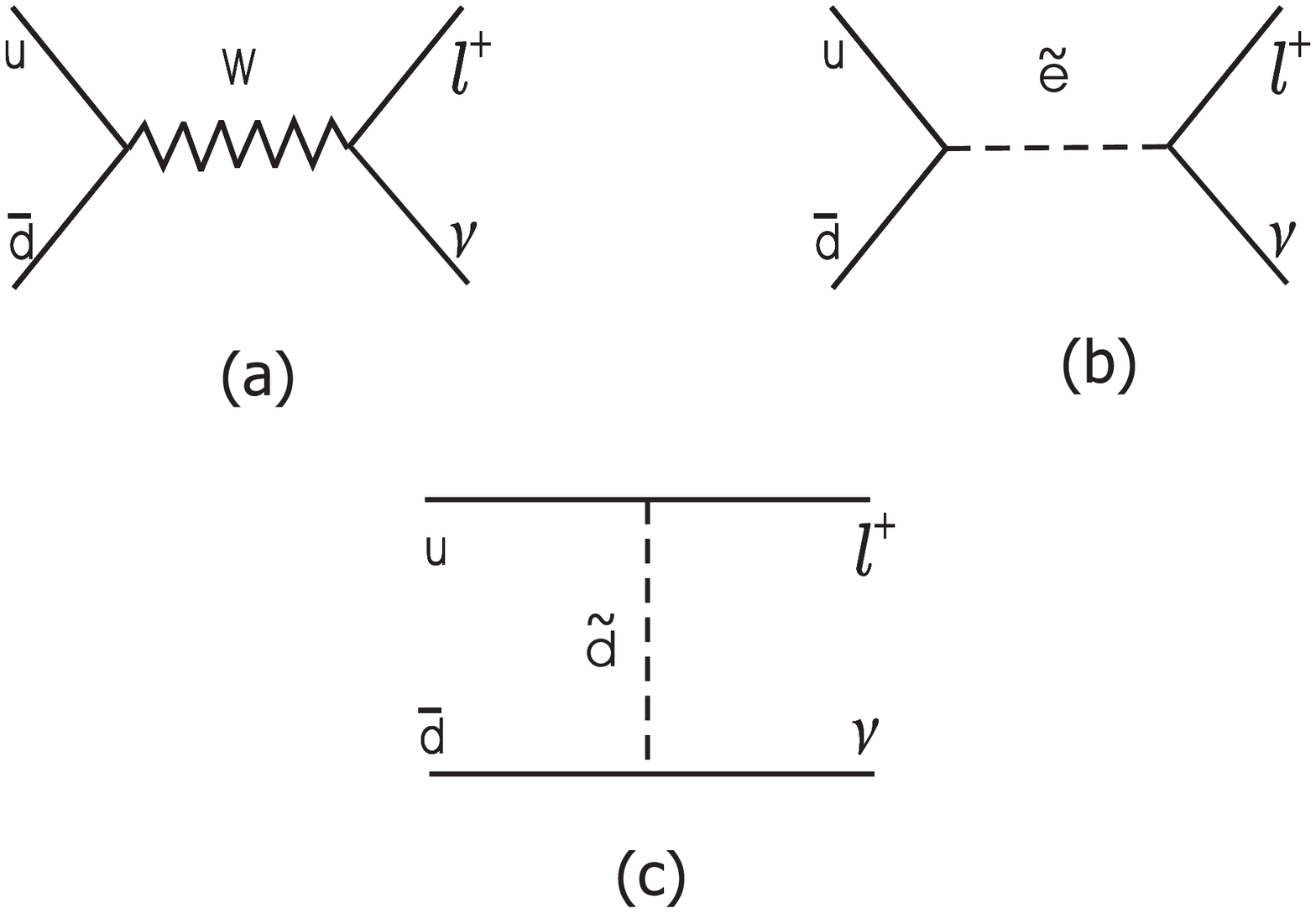}}
\vspace{1.5cm} \caption{ (a) The standard
model W-exchange and (b), (c) the R-parity violating
supersymmetric contributions to the Drell-Yan lepton pair
production. }
\end{figure}
\end{document}